\documentclass[10pt,a4paper,twocolumn]{article}
\usepackage[english]{babel}
\usepackage[left=0.7in,top=0.7in,bottom=0.9in,right=0.5in]{geometry}
\usepackage[svgnames,x11names]{xcolor}
\usepackage{caption}
\usepackage[caption=false]{subfig}
\usepackage{graphicx}
\usepackage{amsmath}
\usepackage{upgreek}
\usepackage{cite}
\usepackage[colorlinks=true
,urlcolor=blue
,anchorcolor=blue
,citecolor=blue
,filecolor=blue
,linkcolor=blue
,menucolor=blue
,linktocpage=true
,pdfproducer=medialab
]{hyperref}
\usepackage{breakurl}
\usepackage[noindentafter]{titlesec}
\titleformat{\section}
  {\normalfont\fontsize{12}{15}\bfseries}{\thesection}{1em}{}

\usepackage{hyphenat}
\newcommand\mcmule{{\sc McMule}}
\begin{document}
\twocolumn[{%
\begin{@twocolumnfalse}
\vspace{-3em}
\begin{flushright}
PSI-PR-21-11\\
ZU-TH 25/21\\
IPPP/20/112
\end{flushright}
\vspace{1em}

\begin{center}
{\Large\bf Bhabha scattering at NNLO with next-to-soft stabilisation}
\\[2em]
{
Pulak Banerjee$^{a}$,
Tim Engel$^{a,b}$,
Nicolas Schalch$^{c}$,
Adrian Signer$^{a,b}$,
Yannick Ulrich$^{d}$
}\\[0.1in]
{\sl ${}^a$ Paul Scherrer Institut,
CH-5232 Villigen PSI, Switzerland \\
${}^b$ Physik-Institut, Universit\"at Z\"urich, 
Winterthurerstrasse 190,
CH-8057 Z\"urich, Switzerland\\
${}^c$ Albert Einstein Center for Fundamental Physics, Institut f\"ur
Theoretische Physik,\\ Universit\"at Bern,
Sidlerstrasse 5, CH-3012 Bern, Switzerland\\
${}^d$ Institute for Particle Physics Phenomenology, University of
Durham, \\
South Road, Durham DH1 3LE, United Kingdom
}
\setcounter{footnote}{0}
\end{center}

\vspace{0.2em}

\begin{center}
\begin{minipage}{6in}
{\small A critical subject in fully differential QED calculations originates from
numerical instabilities due to small fermion masses that act
as regulators of collinear singularities. At next-to-next-to-leading
order (NNLO) a major challenge is therefore to find a stable
implementation of numerically delicate real-virtual matrix elements.
In the case of Bhabha scattering this has so far prevented the
development of a fixed-order Monte Carlo at NNLO accuracy. In this
paper we present a new method for stabilising the real-virtual matrix
element. It is based on the expansion for soft photon energies
including the non-universal subleading term calculated with the method
of regions. We have applied this method to Bhabha scattering to obtain
a stable and efficient implementation within the \mcmule{} framework.
We therefore present for the first time fully differential results for
the photonic NNLO corrections to Bhabha scattering.}

\end{minipage}
\end{center}
\vspace{2em}
\end{@twocolumnfalse}}]

\section{\label{sec:intro}Introduction}

Electron-positron or Bhabha scattering is one of the best studied
processes in the Standard Model~\cite{Actis:2010gg}. It is well suited
for luminosity measurements at $e^+ e^-$ colliders because of its
large cross section and clean signature. Furthermore, for energies well below the electroweak scale the radiative
corrections are dominated by quantum electrodynamics (QED) which
allows for a very precise theory prediction. As a consequence, much
work has been put into the calculation of higher-order matrix elements
as well as the development of Monte-Carlo event generators.

The next-to-leading order (NLO) matrix elements have been known in the
full Standard Model for quite some time~\cite{Consoli:1979xw,
Bohm:1986fg, Berends:1983fs, Caffo:1984jb}. At next-to-next-to-leading
order (NNLO) the situation is different. In the case of the
electroweak corrections only logarithmically enhanced terms have been
calculated~\cite{Kuhn:2001hz, Feucht:2004rp,
Jantzen:2005az, Penin:2011aa}.  On the QED side much more is known.
The full two-loop matrix element with vanishing electron mass was
calculated some time ago~\cite{Bern:2000ie}. Subsequently, this result
was extended to also include leading-order mass
effects~\cite{Penin:2005eh, Mitov:2006xs, Becher:2007cu,
Actis:2007gi}.  The subset of the two-loop matrix element containing
closed electron loops has been computed without any
approximations~\cite{Bonciani:2004gi}. Although the exact mass
dependence of the full two-loop contribution is still not known,
leading power-suppressed mass effects were recently taken into account
in~\cite{Penin:2016wiw}. The one-loop corrections to the radiative matrix element were calculated in~\cite{Actis:2009uq}.

In addition to the work that has been put into the calculation of the
matrix elements various Monte-Carlo event generators were developed, combining
the matrix elements to physical observables such that non-trivial
detector geometries and acceptances can be taken into
account~\cite{Berends:1983fs, Berends:1987jm, Jadach:1995nk,
Placzek:1999xc, Arbuzov:1997pj, CarloniCalame:2000pz,
CarloniCalame:2001ny, CarloniCalame:2003yt, Arbuzov:1999db,
Caffo:1996mi, Jadach:1991cg}. In particular, the {\tt
BABAYAGA} event generator that is based on the matching of the exact
NLO results to a parton shower algorithm has achieved a precision of
below $0.1\%$~\cite{Balossini:2006wc}. A detailed analysis of the impact of fixed-order fermionic NNLO contributions was presented in~\cite{CarloniCalame:2011zq}.

Even though all necessary ingredients are available, a Monte Carlo that
includes also NNLO photonic corrections was missing.  The main
bottleneck in this regard has been the real-virtual contribution that
suffers from numerical instabilities when integrated over the phase
space of the emitted photon. The source of these instabilities can be
traced back to the disparate scales in the process introduced by the
small electron mass that acts as a regulator of collinear divergences.
This problem is exacerbated in the presence of soft radiation. 

In this paper we present a method to reliably integrate the
real-virtual matrix element over the full phase space. It is based on
the expansion for small photon energies
$E_\gamma\equiv\xi\times\sqrt{s}/2$ including the non-universal
next-to-soft contribution at $\mathcal{O}(\xi^{-1})$. To verify our
method we have compared with approximate results from {\tt BABAYAGA}
at the cross section as well as at the differential level and found
agreement within the expected $0.1\%$ precision. With this method it is therefore possible to make
reliable predictions for Bhabha scattering at the differential level
including the full set of NNLO QED corrections. 

This paper is organised as follows: We begin by briefly introducing
our calculational framework in Section~\ref{sec:calculation}. The main
result is presented in Section~\ref{sec:nts} where we describe how the
stabilisation of the real-virtual matrix element was achieved via the
next-to-soft approximation. We verify our method in
Section~\ref{sec:results} and conclude in
Section~\ref{sec:conclusion}.

\section{\label{sec:calculation}Overview of the calculation}

We consider the scattering process
\begin{equation}
e^-(p_1) e^+(p_2) \rightarrow e^-(p_3) e^+(p_4) \{\gamma(p_5) \gamma(p_6)\}
\end{equation}
up to NNLO in QED. As we are mainly interested in establishing the
stabilisation method we restrict ourselves to purely photonic
corrections, i.e. we do not take into account contributions from
closed fermion loops. Ultraviolet (UV) and infrared (IR)
divergences are regularised in $d=4-2\epsilon$ dimensions and the
renormalisation is performed in the on-shell scheme.

All tree-level and one-loop matrix elements were calculated with the
full electron mass dependence. The corresponding diagrams were
generated using QGraf~\cite{NOGUEIRA1993279} and evaluated with the Mathematica code
Package-X~\cite{Patel:2015tea}. In the case of the numerically
delicate real-virtual matrix element this Mathematica calculation
serves mostly as a reference calculation. In the bulk of the phase
space we instead rely on OpenLoops~\cite{Buccioni:2017yxi,Buccioni:2019sur}. As we will
discuss in Section~\ref{sec:nts}, for small photon energies we switch to a next-to-soft approximation.

As mentioned in the introduction the full mass dependence of the
photonic two-loop matrix element is not known. However, for most
practical applications we can assume the scale hierarchy $m^2 \ll
Q^2\in\{s,t,u\}$ with the electron mass $m$ and the Mandelstam
invariants $s=(p_1+p_2)^2$, $t=(p_1-p_3)^2$, and $u=(p_1-p_4)^2$. For
our purposes it is therefore justified to neglect the power-suppressed
terms of $\mathcal{O}(m^2/Q^2)$. Due to the universal structure of
collinear divergences the leading mass effects can be straightforwardly included based on
the massless result. This massification procedure was developed in the
context of Bhabha scattering~\cite{Penin:2005eh, Mitov:2006xs,
Becher:2007cu} and was recently extended to processes with a heavy
mass~\cite{Engel:2018fsb}.

The matrix elements are implemented in the integrator \mcmule{}, a Monte Carlo for MUons and other LEptons~\cite{Banerjee:2020rww}. This framework is based on the FKS$^\ell$
subtraction scheme~\cite{Engel:2019nfw} which is an extension of the
original FKS scheme~\cite{Frixione:1995ms, Frederix:2009Yq} beyond NLO
for QED.  This subtraction scheme allows to consistently remove the
singularities arising from soft photon emission in order to calculate
observables in a fully differential way. The simplicity of the
FKS$^\ell$ subtraction scheme is due to the absence of collinear and
the simple structure of soft singularities. All collinear divergences
are regulated by finite fermion masses. Following the notation
from~\cite{Engel:2019nfw} the soft singularities exponentiate
according to the YFS formula~\cite{YENNIE1961379}
\begin{equation}\label{eq:yfs}
    \sum_{l=0}^\infty \mathcal{M}_n^{(\ell)} = e^{-\hat{\mathcal{E}}} \sum_{l=0}^\infty \mathcal{M}_n^{(\ell) f}.
\end{equation}
All soft poles of the $\ell$-loop matrix element (squared amplitude)
with $n$ final-state particles $\mathcal{M}_n^{(\ell)}$ are absorbed
in the universal integrated eikonal factor $\hat{\mathcal{E}}$,
rendering $\mathcal{M}_n^{(\ell) f}$ finite. This formula can be seen
as a consequence of the universal behaviour of radiative matrix
elements in the soft limit
\begin{equation}\label{eq:softlimit}
    \lim_{\xi \rightarrow 0} \xi^2 \mathcal{M}_{n+1}^{(\ell)}
    =\mathcal{E} \mathcal{M}_{n}^{(\ell)},
\end{equation}
with the scaled photon energy $\xi = 2 E_\gamma/\sqrt{s}$ and the
eikonal factor $\mathcal{E}$.

The most challenging part of the calculation presented here is to
ensure a reliable integration of the real emission contributions in
the phase-space region where the photon becomes collinear to the
emitting fermion or where it becomes soft. In the former case the
smallness of the electron mass acting as an infrared regulator results
in large pseudo-collinear singularities. To address this issue we use
a dedicated tuning of the phase-space parametrisation to help the
{\tt{vegas}} integration~\cite{Lepage:1980jk} find and deal with these
problematic regions. In the latter case the integrand develops an
unregularised soft singularity that is subtracted with the IR
counterterm, resulting in a large cancellation. For this cancellation
to work the matrix element has to be evaluated with very high
precision. Due to the analytical and algebraic complexity of the
real-virtual matrix element this is a highly non-trivial task and has
presented the main obstacle to a complete, fully differential NNLO
calculation of Bhabha scattering in the past. Our solution to this
problem is the main result of this paper and is discussed in detail in
the next section.

\section{\label{sec:nts}Real-virtual stabilisation via next-to-soft approximation}

\begin{figure}[t]
\centering
\subfloat[Arbitrary phase-space point]{
         \includegraphics[width=.465\textwidth]{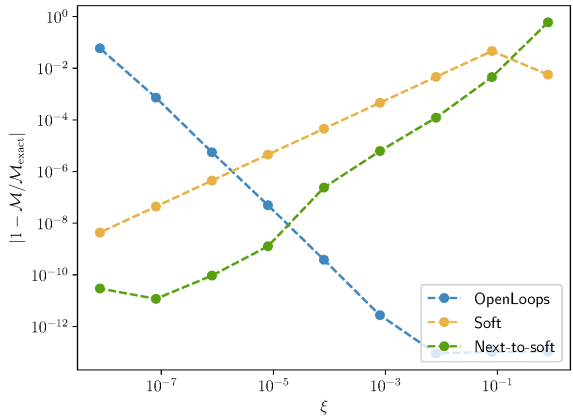}
         \label{fig:softlimit_arbitrary}
}\\
\subfloat[Initial-state collinear phase-space point]{
         \includegraphics[width=.465\textwidth]{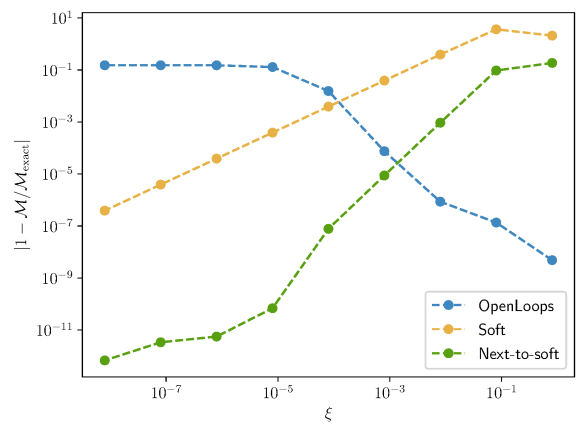}
         \label{fig:softlimit_collinear}
}
\captionof{figure}{\label{fig:softlimit}
Behaviour of the soft approximations and of OpenLoops compared
to the `exact' real-virtual matrix element in the soft limit $E_\gamma =
\xi\times \sqrt{s}/2\rightarrow 0$.}
\end{figure}

This section discusses how an implementation of the real-virtual
matrix element can be obtained that ensures a stable and efficient
integration in the soft phase-space region. As alluded to above, any
general-purpose calculation of a one-loop matrix element will run into
numerical instabilities at some point. In particular, for processes
with an external photon with ever smaller energies, the IR-subtracted
matrix element is a typical numerical pitfall whereby two expressions
diverging as $1/\xi$ are combined to obtain an integrable integrand
diverging as $1/\sqrt{\xi}$. The crucial question is whether these
instabilities appear only for small enough $\xi$ such that the
integration can be done reliably. In this context, QED calculations
are particularly delicate since the final states tend to be much less
inclusive than jet cross sections computed for hadronic collisions.

We use OpenLoops~\cite{Buccioni:2019sur} for the bulk of the phase space since it shows a
remarkable numerical stability. In order to test for which values of
$\xi$ the instabilities start to appear, we compare OpenLoops to a
dedicated computation of the real-virtual matrix element in
Mathematica using arbitrary precision arithmetic. In
Figure~\ref{fig:softlimit} we show the deviation of OpenLoops from the
`exact' Mathematica result. For illustration we use an arbitrary phase-space point as
well as one where the photon is emitted nearly collinear to the
initial-state electron. For the former, at $\xi=10^{-5}$ the relative
error is $10^{-8}$. In the collinear case the numerical instabilities
are strongly enhanced with a relative difference of $10^{-1}$ for
$\xi=10^{-5}$. All numbers that enter Figure~\ref{fig:softlimit} including the particle momenta $p_i$ of the two phase-space points are publicly available under~\cite{website}.

An obvious idea is to expand the real-virtual matrix element for small
photon energies and to switch to this approximation for sufficiently
small $\xi$. The leading $\mathcal{O}(\xi^{-2})$ term in this
expansion is given by~\eqref{eq:softlimit} and can be easily
calculated based on the one-loop matrix element $\mathcal{M}_n^{(1)}$.
Using this approximation amounts to using the same algebraic
expression for both terms in the subtracted integrand, albeit with
different kinematics. As can be seen from Figure~\ref{fig:softlimit}
this approach is insufficient in the collinear region. If an accuracy
below $10^{-3}$ is to be aimed at, in this case one has to switch to the expansion
at $\xi\sim10^{-3}$. However, the exact matrix element is
not sufficiently well approximated by the leading soft contribution in
this region. To ensure a decent approximation we have therefore to
include the non-universal $\mathcal{O}(\xi^{-1})$ term in the soft
expansion.

The next-to-soft terms of tree-level matrix elements have been
considered a long time ago (Low-Burnett-Kroll theorem)~\cite{Low:1958sn,Burnett:1967km}. Going
beyond tree level, it is tempting to try to apply
effective-field-theory methods. However, for QED with massive fermions
the appropriate effective theory is the QED version of heavy quark
effective theory and the genuine one-loop contribution to the next-to-soft effects is expected
to be given by the process dependent soft function. From a practicable point of view we
have thus decided to directly calculate the non-universal
$\mathcal{O}(\xi^{-1})$ term in the soft expansion.

To be precise, we have computed the real-virtual matrix element in
terms of scalar Passarino-Veltman functions using the
Mathematica calculation of the real-virtual matrix element described
in the previous section. Next, we have employed the power counting 
\begin{align}
\begin{split}
    &p_i \rightarrow p_i \quad \text{for} \quad i\in\{1,2,3,4\}, \\
    &p_5 \rightarrow \lambda p_5, \\
    &m \rightarrow m,
\end{split}
\end{align}
and expanded in the book-keeping parameter $\lambda$. The expansion of
the rational coefficients and simple Passarino-Veltman
functions was performed with Mathematica. More complicated triangle-
and box-functions were expanded at the (loop-)integrand level using
the method of regions~\cite{Beneke:1997zp}. For the purpose of
calculational efficiency we have used its formulation in the
parametric representation~\cite{Smirnov:1999bza}. In this case, the
contributing regions can be easily found using the public code {\tt
asy.m}~\cite{Jantzen:2012mw}. Most resulting
integrals could be straightforwardly computed. The remaining ones
were calculated using Mellin-Barnes techniques~\cite{Smirnov:1999gc,
Tausk:1999vh}. For the most involved integrals a two-fold
Mellin-Barnes representation was necessary.  However, the reduction to
single contour integrals was possible in this case by resolving the
singularity structure with the Mathematica package {\tt
MBresolve.m}~\cite{Smirnov:2009up}. In summary, the main technical
difficulties in performing the next-to-soft expansion are the
calculation of the integrals and the treatment of large intermediate
expressions.

We have checked that the first term of the expansion indeed reproduces
the result from~\eqref{eq:softlimit}. The non-universal subleading
contribution was verified numerically. This is also shown in
Figure~\ref{fig:softlimit} where the inclusion of the next-to-soft
$\mathcal{O}(\xi^{-1})$ term significantly improves the approximation.
This allows us to switch to a reliable expansion as early as $\xi \sim
10^{-3}$. We can therefore conclude that the next-to-soft approach
ensures the numerical stability of the real-virtual matrix element for
small photon energies which is a prerequisite for the IR subtraction
to work. This is further emphasised by comparing integrated results with and without stabilisation. While the next-to-soft stabilisation ensures that results after successive Monte Carlo iterations are in agreement with each other, a drifting mean value is observed otherwise resulting in a significant discrepancy between the two results. Furthermore, the evaluation of the obtained expansion is
a few $100$ to over a $1000$ times faster than OpenLoops, depending
on the details of the kinematics. Since {\tt{vegas}} tends to
sample predominantly in the soft and collinear region this speed-up is
noticeable even in the integration over the full phase space. While OpenLoops provides settings to work at higher accuracy this
comes at a cost of speed.

\section{\label{sec:results}Results and verification}

To test the next-to-soft approach of stabilising the real-virtual
contribution we have compared to {\tt{BABAYAGA}}. For all results
presented in this section we have switched from OpenLoops to the
next-to-soft approximation at $\xi=10^{-3}$. As mentioned in the
introduction, the event generator {\tt{BABAYAGA}} is based on a
parton shower algorithm matched to the exact NLO result. Contrary to
our complete fixed-order calculation it therefore only gives the
logarithmically enhanced contributions at NNLO. For the comparison we
use set-up (a) of~\cite{Balossini:2006wc} that is tailored to $\phi$
factories with a centre-of-mass energy of $\sqrt{s} = 1020~\rm{MeV}$.
The detector configuration is approximated with the kinematical cuts
\begin{align}
\begin{split}
    &E_{\rm{min}} = 408~\rm{MeV}, \\
    &20^0 < \theta_\pm < 160^0, \\
    &\zeta_{\rm{max}} = 10^0,
\end{split}
\end{align}
where $E_{\rm min}$ is the minimum energy of the final-state
electron/positron, $\theta_{-}(\theta_{+})$ is the scattering angle in
the centre-of-mass frame between the incoming and outgoing electron
(positron), and $\zeta_{\rm max}$ is the maximally allowed
acollinearity $\zeta=|180^0-\theta_{+}-\theta_{-}|$.

\begin{figure}[t]
\centering
 \begin{tabular}{c | r c || c r} 
 \phantom{largerrr} & \multicolumn{2}{c||}{$\sigma/\rm \upmu b$} & \multicolumn{2}{c}{\hspace{0.3cm}$\delta K^{(i)}/\%$\hspace{0.3cm}} \\
  &  \multicolumn{1}{c}{\hspace{0.3cm}\tt McMule\hspace{0.3cm}}  & \multicolumn{1}{c||}{\hspace{0.3cm} \tt BABAYAGA \hspace{0.3cm}}  & \multicolumn{1}{c}{\hspace{0.3cm} \tt McMule \hspace{0.3cm}} \\ [0.5ex] 
  \hline
 \rule{0pt}{3ex}
 $\sigma^{(0)}$ & \tt  6.8557  & \hspace{0.3cm}\tt 6.8557  & & \\[0.5ex]
 \hline
 \rule{0pt}{3ex}
 $\sigma^{(1)}$ & \tt  -0.7957  &\hspace{0.1cm}\tt -0.7957& \tt -11.606 \\[0.5ex]
 \hline
 \rule{0pt}{3ex}
 $\sigma^{(2)}$ & \tt    0.0312  &\hspace{0.3cm}\tt    0.0267 & \hspace{0.35cm}\tt   0.515 \\[0.5ex]
 \hline
 \hline
 \rule{0pt}{2.7ex}
 $\sigma_{2}$   & \tt  6.0912  &\hspace{0.3cm}\tt  6.0868 & & \\[0.5ex]
\end{tabular}
\captionof{table}{\label{tab:xsection}
Comparison of our exact fixed-order calculation for the total cross
section with the full LO and NLO as well as the approximate NNLO
results from {\tt BABAYAGA}~\cite{Carlo}. All digits given are
significant compared to the error of the numerical integration.}

\centering
\includegraphics[width=0.48\textwidth]{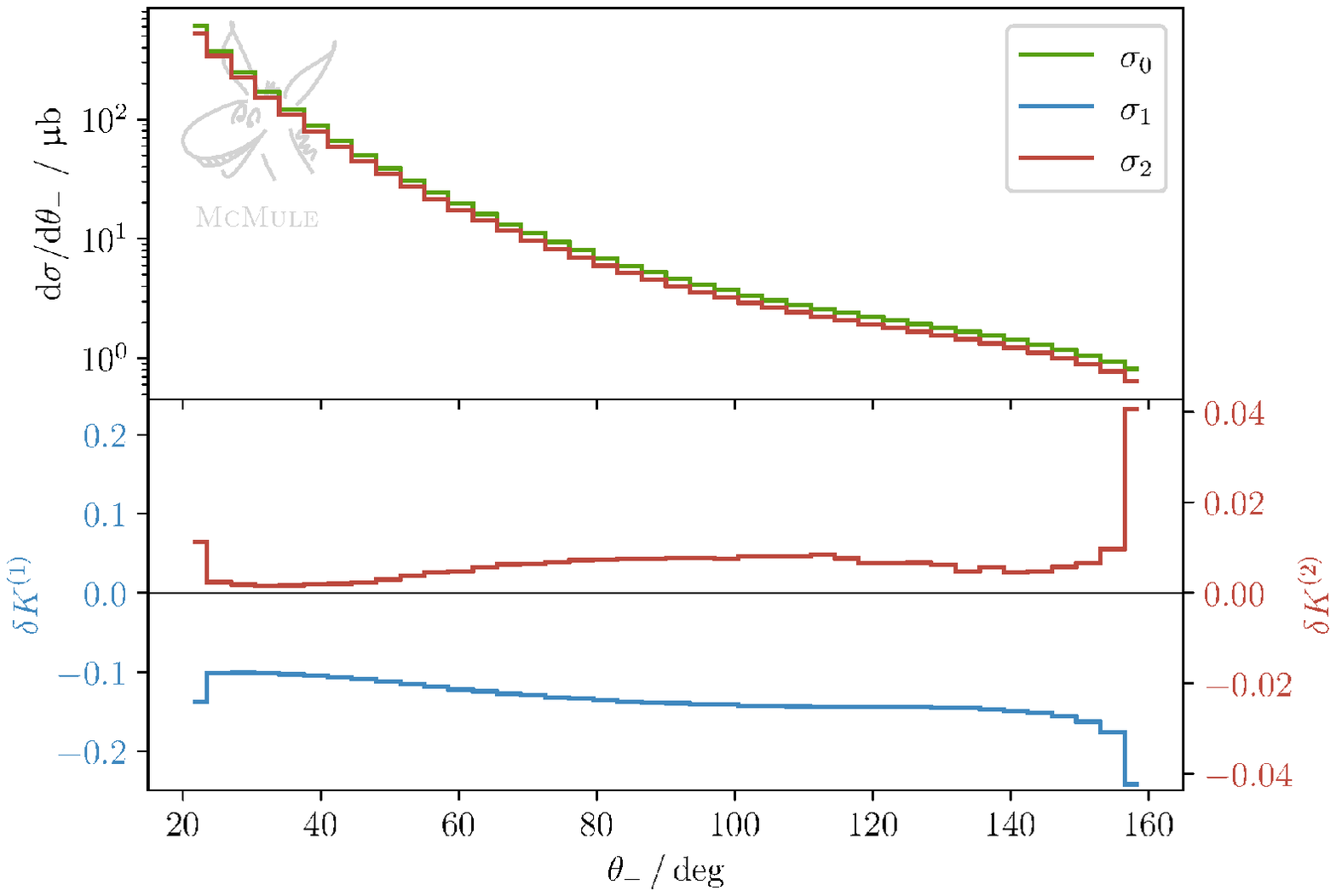}
\captionof{figure}{\label{fig:distribution}
The differential cross section w.r.t. $\theta_{-}$ at LO (green) and
NNLO (red). The NLO and NNLO $K$ factors are shown in blue and red,
respectively.}
\end{figure}

The order-by-order contributions, $\sigma^{(i)}$, to the integrated
cross section, $\sigma_2=\sigma^{(0)} + \sigma^{(1)} + \sigma^{(2)}$,
are presented in Table~\ref{tab:xsection}. Additionally, we show the
corresponding $K$ factors defined as
\begin{equation}
    K^{(i)}=1+\delta K^{(i)} =
    \frac{\sigma_{i}}{\sigma_{i-1}}.
\end{equation}
To avoid comparing to contributions from the parton shower beyond NNLO
we have not directly compared to~\cite{Balossini:2006wc} but instead
were provided truncated results~\cite{Carlo}. We find complete
agreement for the LO as well as the NLO result. Our fixed-order NNLO
correction $\sigma^{(2)}$ agrees at the level of $17\%$ with the NNLO
contribution from the matched parton shower. This translates to an
agreement for the total cross section $\sigma_2$ of $0.07\%$,
consistent with the $0.1\%$ precision aimed at
in~\cite{Balossini:2006wc}. We note that the sizable $K$ factors are a
consequence of the cut on the acollinearity that suppresses hard
radiation.

Figure~\ref{fig:distribution} shows differential results with respect
to the electron scattering angle $\theta_{-}$. The differential cross
section at LO as well as at NNLO are displayed in the upper panel. In
addition, the lower panel shows the differential $K$ factors
\begin{equation}
   K^{(i)} = 1+\delta K^{(i)}=
   \frac{d\sigma_{i}/d \theta_{-}}{d\sigma_{i-1}/d \theta_{-}}.
\end{equation}
The comparison with the truncated parton shower at the differential
level yields a similar result as for the total cross section, i.e.
complete agreement up to NLO and deviations of below $0.1\%$ at NNLO.

\section{\label{sec:conclusion}Conclusion and outlook}

One of the main complications in the calculation of fully differential
higher-order corrections in QED is the occurrence of numerical
instabilities in real-emission contributions due to finite but small fermion
masses acting as collinear regulators. In the case of Bhabha
scattering instabilities arising in the real-virtual matrix element
have represented the main bottleneck for a fixed-order Monte Carlo at
NNLO accuracy. 

In this paper we have presented a new method that ensures a stable and
efficient integration of these problematic contributions. Since the
main instabilities occur for soft photon emission we have expanded the
real-virtual matrix element for small photon energies including the
non-universal subleading contribution using the method of regions.
This then allows for the reliable use of OpenLoops in the bulk of the
phase space and to switch to the next-to-soft approximation otherwise.
While an analogous approach could also be used for the real and the
double-real matrix element, a stable and fast implementation can be
obtained without the next-to-soft stabilisation.

We were therefore able to calculate for the first time the photonic
NNLO corrections for Bhabha scattering in a fully differential way.
This was implemented in the \mcmule{} framework. We have cross-checked
our exact NNLO results at the level of the total cross section as well
as for differential distributions with the logarithmic approximation
implemented in the parton shower generator {\tt BABAYAGA}.

Numerical instabilities of the kind described above are a critical
point of higher-order QED calculations. We therefore expect that the
next-to-soft method will prove useful in other processes as well. This
is obvious in the case of M\o ller scattering which is related to the
Bhabha process via crossing. Corresponding results relevant for the experiment PRad~II~\cite{Gasparian:2020hog} will be presented in a
forthcoming paper~\cite{Banerjee:2021qvi}. Furthermore, in the context of
muon-electron scattering our approach could turn out to be invaluable
where a fully differential NNLO calculation is highly
desirable~\cite{Abbiendi:2016xup} and therefore aimed
at~\cite{Banerjee:2020tdt}.

Because of the wide range of applicability of the next-to-soft
expansion, an investigation of a potential universal structure would
be desirable to allow for a more efficient calculation of the
expansion. This could be done in the framework of heavy quark
effective theory. Furthermore, a similar approach could be
pursued in the collinear region. Switching to a leading collinear
expansion could result in a significant speed-up of the phase-space
integration. This would entail the calculation of the currently
unknown one-loop splitting functions for massive fermions.

\subsection*{Acknowledgement} 
It is a great pleasure to thank C.~Carloni Calame for engaging in a
detailed comparison with {\tt BABAYAGA}. We are also grateful to M.~Zoller for providing the real-virtual matrix element in OpenLoops as
well as for assisting us with its proper usage. Without the impressive
numerical stability of OpenLoops this project would not have been
possible. Our thanks are further extended to L.~Dixon for sharing the
massless two-loop matrix element in electronic form. Finally, we would
also like to thank T.~Becher for helpful discussions.  PB and TE
acknowledge support by the Swiss National Science Foundation (SNF)
under contract 200021\_178967.  YU acknowledges partial support by a
Forschungskredit of the University of Zurich under contract number
FK-19-087 as well as by the UK Science and Technology Facilities
Council (STFC) under grant ST/T001011/1. PB acknowledges support by
the European Union's Horizon 2020 research and innovation program
under the Marie Sk\l{}odowska-Curie grant agreement No 701647.

\bibliographystyle{JHEP}
\small\bibliography{bhabha}

\providecommand{\noopsort}[1]{}\providecommand{\singleletter}[1]{#1}%

\providecommand{\href}[2]{#2}\begingroup\raggedright\begin{thebibliography}{10}

\bibitem{Actis:2010gg}
{\scshape Working Group on Radiative Corrections, Monte Carlo Generators for
  Low Energies} collaboration, S.~Actis et~al., \emph{{Quest for precision in
  hadronic cross sections at low energy: Monte Carlo tools vs. experimental
  data}}, \href{https://doi.org/10.1140/epjc/s10052-010-1251-4}{\emph{Eur.
  Phys. J. C} {\bfseries 66} (2010) 585}
  [\href{https://arxiv.org/abs/0912.0749}{{\ttfamily 0912.0749}}].

\bibitem{Consoli:1979xw}
M.~Consoli, \emph{{One Loop Corrections to $e^+ e^- \to e^+ e^-$ in the
  Weinberg Model}},
  \href{https://doi.org/10.1016/0550-3213(79)90235-9}{\emph{Nucl. Phys. B}
  {\bfseries 160} (1979) 208}.

\bibitem{Bohm:1986fg}
M.~Bohm, A.~Denner and W.~Hollik, \emph{{Radiative Corrections to Bhabha
  Scattering at High-Energies. 1. Virtual and Soft Photon Corrections}},
  \href{https://doi.org/10.1016/0550-3213(88)90650-5}{\emph{Nucl. Phys. B}
  {\bfseries 304} (1988) 687}.

\bibitem{Berends:1983fs}
F.~A. Berends and R.~Kleiss, \emph{{Distributions in the Process $e^+ e^-
  \rightarrow e^+ e^- (\gamma)$}},
  \href{https://doi.org/10.1016/0550-3213(83)90558-8}{\emph{Nucl. Phys. B}
  {\bfseries 228} (1983) 537}.

\bibitem{Caffo:1984jb}
M.~Caffo, R.~Gatto and E.~Remiddi, \emph{{Hard collinear photons, high-energy
  radiative corrections to Bhabha scattering}},
  \href{https://doi.org/10.1016/0550-3213(85)90453-5}{\emph{Nucl. Phys. B}
  {\bfseries 252} (1985) 378}.

\bibitem{Kuhn:2001hz}
J.~H. Kuhn, S.~Moch, A.~A. Penin and V.~A. Smirnov,
  \emph{{Next-to-next-to-leading logarithms in four fermion electroweak
  processes at high-energy}},
  \href{https://doi.org/10.1016/S0550-3213(01)00454-0}{\emph{Nucl. Phys. B}
  {\bfseries 616} (2001) 286}
  [\href{https://arxiv.org/abs/hep-ph/0106298}{{\ttfamily hep-ph/0106298}}].

\bibitem{Feucht:2004rp}
B.~Feucht, J.~H. Kuhn, A.~A. Penin and V.~A. Smirnov, \emph{{Two loop Sudakov
  form-factor in a theory with mass gap}},
  \href{https://doi.org/10.1103/PhysRevLett.93.101802}{\emph{Phys. Rev. Lett.}
  {\bfseries 93} (2004) 101802}
  [\href{https://arxiv.org/abs/hep-ph/0404082}{{\ttfamily hep-ph/0404082}}].

\bibitem{Jantzen:2005az}
B.~Jantzen, J.~H. Kuhn, A.~A. Penin and V.~A. Smirnov, \emph{{Two-loop
  electroweak logarithms in four-fermion processes at high energy}},
  \href{https://doi.org/10.1016/j.nuclphysb.2005.10.010}{\emph{Nucl. Phys. B}
  {\bfseries 731} (2005) 188}
  [\href{https://arxiv.org/abs/hep-ph/0509157}{{\ttfamily hep-ph/0509157}}].

\bibitem{Penin:2011aa}
A.~A. Penin and G.~Ryan, \emph{{Two-loop electroweak corrections to high energy
  large-angle Bhabha scattering}},
  \href{https://doi.org/10.1007/JHEP11(2011)081}{\emph{JHEP} {\bfseries 11}
  (2011) 081} [\href{https://arxiv.org/abs/1112.2171}{{\ttfamily 1112.2171}}].

\bibitem{Bern:2000ie}
Z.~Bern, L.~J. Dixon and A.~Ghinculov, \emph{{Two loop correction to Bhabha
  scattering}}, \href{https://doi.org/10.1103/PhysRevD.63.053007}{\emph{Phys.
  Rev. D} {\bfseries 63} (2001) 053007}
  [\href{https://arxiv.org/abs/hep-ph/0010075}{{\ttfamily hep-ph/0010075}}].

\bibitem{Penin:2005eh}
A.~A. Penin, \emph{{Two-loop photonic corrections to massive Bhabha
  scattering}},
  \href{https://doi.org/10.1016/j.nuclphysb.2005.11.016}{\emph{Nucl. Phys. B}
  {\bfseries 734} (2006) 185}
  [\href{https://arxiv.org/abs/hep-ph/0508127}{{\ttfamily hep-ph/0508127}}].

\bibitem{Mitov:2006xs}
A.~Mitov and S.~Moch, \emph{{The Singular behavior of massive QCD amplitudes}},
  \href{https://doi.org/10.1088/1126-6708/2007/05/001}{\emph{JHEP} {\bfseries
  05} (2007) 001} [\href{https://arxiv.org/abs/hep-ph/0612149}{{\ttfamily
  hep-ph/0612149}}].

\bibitem{Becher:2007cu}
T.~Becher and K.~Melnikov, \emph{{Two-loop QED corrections to Bhabha
  scattering}},
  \href{https://doi.org/10.1088/1126-6708/2007/06/084}{\emph{JHEP} {\bfseries
  06} (2007) 084} [\href{https://arxiv.org/abs/0704.3582}{{\ttfamily
  0704.3582}}].

\bibitem{Actis:2007gi}
S.~Actis, M.~Czakon, J.~Gluza and T.~Riemann, \emph{{Two-loop fermionic
  corrections to massive Bhabha scattering}},
  \href{https://doi.org/10.1016/j.nuclphysb.2007.06.023}{\emph{Nucl. Phys. B}
  {\bfseries 786} (2007) 26} [\href{https://arxiv.org/abs/0704.2400}{{\ttfamily
  0704.2400}}].

\bibitem{Bonciani:2004gi}
R.~Bonciani, A.~Ferroglia, P.~Mastrolia, E.~Remiddi and J.~J. van~der Bij,
  \emph{{Two-loop $N_F=1$ QED Bhabha scattering differential cross section}},
  \href{https://doi.org/10.1016/j.nuclphysb.2004.09.015}{\emph{Nucl. Phys. B}
  {\bfseries 701} (2004) 121}
  [\href{https://arxiv.org/abs/hep-ph/0405275}{{\ttfamily hep-ph/0405275}}].

\bibitem{Penin:2016wiw}
A.~A. Penin and N.~Zerf, \emph{{Two-loop Bhabha Scattering at High Energy
  beyond Leading Power Approximation}},
  \href{https://doi.org/10.1016/j.physletb.2016.07.077}{\emph{Phys. Lett. B}
  {\bfseries 760} (2016) 816}
  [\href{https://arxiv.org/abs/1606.06344}{{\ttfamily 1606.06344}}].

\bibitem{Actis:2009uq}
S.~Actis, P.~Mastrolia and G.~Ossola, \emph{{NLO QED Corrections to
  Hard-Bremsstrahlung Emission in Bhabha Scattering}},
  \href{https://doi.org/10.1016/j.physletb.2009.11.035}{\emph{Phys. Lett. B}
  {\bfseries 682} (2010) 419}
  [\href{https://arxiv.org/abs/0909.1750}{{\ttfamily 0909.1750}}].

\bibitem{Berends:1987jm}
F.~A. Berends, R.~Kleiss and W.~Hollik, \emph{{Radiative Corrections to Bhabha
  Scattering at High-Energies. 2. Hard Photon Corrections and Monte Carlo
  Treatment}}, \href{https://doi.org/10.1016/0550-3213(88)90651-7}{\emph{Nucl.
  Phys. B} {\bfseries 304} (1988) 712}.

\bibitem{Jadach:1995nk}
S.~Jadach, W.~Placzek and B.~F.~L. Ward, \emph{{BHWIDE 1.00:
  $\mathcal{O}(\alpha)$ YFS exponentiated Monte Carlo for Bhabha scattering at
  wide angles for LEP-1 / SLC and LEP-2}},
  \href{https://doi.org/10.1016/S0370-2693(96)01382-2}{\emph{Phys. Lett. B}
  {\bfseries 390} (1997) 298}
  [\href{https://arxiv.org/abs/hep-ph/9608412}{{\ttfamily hep-ph/9608412}}].

\bibitem{Placzek:1999xc}
W.~Placzek, S.~Jadach, M.~Melles, B.~F.~L. Ward and S.~A. Yost,
  \emph{{Precision calculation of Bhabha scattering at LEP}},  in \emph{{4th
  International Symposium on Radiative Corrections: Applications of Quantum
  Field Theory to Phenomenology}}, 1, 1999,
  \href{https://arxiv.org/abs/hep-ph/9903381}{{\ttfamily hep-ph/9903381}}.

\bibitem{Arbuzov:1997pj}
A.~B. Arbuzov, G.~V. Fedotovich, E.~A. Kuraev, N.~P. Merenkov, V.~D. Rushai and
  L.~Trentadue, \emph{{Large angle QED processes at $e^+ e^-$ colliders at
  energies below 3-GeV}},
  \href{https://doi.org/10.1088/1126-6708/1997/10/001}{\emph{JHEP} {\bfseries
  10} (1997) 001} [\href{https://arxiv.org/abs/hep-ph/9702262}{{\ttfamily
  hep-ph/9702262}}].

\bibitem{CarloniCalame:2000pz}
C.~M. Carloni~Calame, C.~Lunardini, G.~Montagna, O.~Nicrosini and F.~Piccinini,
  \emph{{Large angle Bhabha scattering and luminosity at flavor factories}},
  \href{https://doi.org/10.1016/S0550-3213(00)00356-4}{\emph{Nucl. Phys. B}
  {\bfseries 584} (2000) 459}
  [\href{https://arxiv.org/abs/hep-ph/0003268}{{\ttfamily hep-ph/0003268}}].

\bibitem{CarloniCalame:2001ny}
C.~M. Carloni~Calame, \emph{{An improved Parton Shower algorithm in QED}},
  \href{https://doi.org/10.1016/S0370-2693(01)01108-X}{\emph{Phys. Lett. B}
  {\bfseries 520} (2001) 16}
  [\href{https://arxiv.org/abs/hep-ph/0103117}{{\ttfamily hep-ph/0103117}}].

\bibitem{CarloniCalame:2003yt}
C.~M. Carloni~Calame, G.~Montagna, O.~Nicrosini and F.~Piccinini, \emph{{The
  BABAYAGA event generator}},
  \href{https://doi.org/10.1016/j.nuclphysbps.2004.02.008}{\emph{Nucl. Phys. B
  Proc. Suppl.} {\bfseries 131} (2004) 48}
  [\href{https://arxiv.org/abs/hep-ph/0312014}{{\ttfamily hep-ph/0312014}}].

\bibitem{Arbuzov:1999db}
A.~B. Arbuzov, \emph{{LABSMC: Monte Carlo event generator for large angle
  Bhabha scattering}},  \href{https://arxiv.org/abs/hep-ph/9907298}{{\ttfamily
  hep-ph/9907298}}.

\bibitem{Caffo:1996mi}
M.~Caffo and H.~Czyz, \emph{{BHAGEN-1PH: A Monte Carlo event generator for
  radiative Bhabha scattering}},
  \href{https://doi.org/10.1016/S0010-4655(96)00134-8}{\emph{Comput. Phys.
  Commun.} {\bfseries 100} (1997) 99}
  [\href{https://arxiv.org/abs/hep-ph/9607357}{{\ttfamily hep-ph/9607357}}].

\bibitem{Jadach:1991cg}
S.~Jadach, E.~Richter-Was, B.~F.~L. Ward and Z.~Was, \emph{{QED multi - photon
  corrections to Bhabha scattering at low angles: Monte Carlo solution}},
  \href{https://doi.org/10.1016/0370-2693(91)90813-6}{\emph{Phys. Lett. B}
  {\bfseries 268} (1991) 253}.

\bibitem{Balossini:2006wc}
G.~Balossini, C.~M. Carloni~Calame, G.~Montagna, O.~Nicrosini and F.~Piccinini,
  \emph{{Matching perturbative and parton shower corrections to Bhabha process
  at flavour factories}},
  \href{https://doi.org/10.1016/j.nuclphysb.2006.09.022}{\emph{Nucl. Phys. B}
  {\bfseries 758} (2006) 227}
  [\href{https://arxiv.org/abs/hep-ph/0607181}{{\ttfamily hep-ph/0607181}}].

\bibitem{CarloniCalame:2011zq}
C.~Carloni~Calame, H.~Czyz, J.~Gluza, M.~Gunia, G.~Montagna, O.~Nicrosini
  et~al., \emph{{NNLO leptonic and hadronic corrections to Bhabha scattering
  and luminosity monitoring at meson factories}},
  \href{https://doi.org/10.1007/JHEP07(2011)126}{\emph{JHEP} {\bfseries 07}
  (2011) 126} [\href{https://arxiv.org/abs/1106.3178}{{\ttfamily 1106.3178}}].

\bibitem{NOGUEIRA1993279}
P.~Nogueira, \emph{{Automatic Feynman Graph Generation}},
  \href{https://doi.org/https://doi.org/10.1006/jcph.1993.1074}{\emph{Journal
  of Computational Physics} {\bfseries 105} (1993) 279}.

\bibitem{Patel:2015tea}
H.~H. Patel, \emph{{Package-X: A Mathematica package for the analytic
  calculation of one-loop integrals}},
  \href{https://doi.org/10.1016/j.cpc.2015.08.017}{\emph{Comput. Phys. Commun.}
  {\bfseries 197} (2015) 276}
  [\href{https://arxiv.org/abs/1503.01469}{{\ttfamily 1503.01469}}].

\bibitem{Buccioni:2017yxi}
F.~Buccioni, S.~Pozzorini and M.~Zoller, \emph{{On-the-fly reduction of open
  loops}}, \href{https://doi.org/10.1140/epjc/s10052-018-5562-1}{\emph{Eur.
  Phys. J. C} {\bfseries 78} (2018) 70}
  [\href{https://arxiv.org/abs/1710.11452}{{\ttfamily 1710.11452}}].

\bibitem{Buccioni:2019sur}
F.~Buccioni, J.-N. Lang, J.~M. Lindert, P.~Maierh\"ofer, S.~Pozzorini, H.~Zhang
  et~al., \emph{{OpenLoops 2}},
  \href{https://doi.org/10.1140/epjc/s10052-019-7306-2}{\emph{Eur. Phys. J. C}
  {\bfseries 79} (2019) 866}
  [\href{https://arxiv.org/abs/1907.13071}{{\ttfamily 1907.13071}}].

\bibitem{Engel:2018fsb}
T.~Engel, C.~Gnendiger, A.~Signer and Y.~Ulrich, \emph{{Small-mass effects in
  heavy-to-light form factors}},
  \href{https://doi.org/10.1007/JHEP02(2019)118}{\emph{JHEP} {\bfseries 02}
  (2019) 118} [\href{https://arxiv.org/abs/1811.06461}{{\ttfamily
  1811.06461}}].

\bibitem{Banerjee:2020rww}
P.~Banerjee, T.~Engel, A.~Signer and Y.~Ulrich, \emph{{QED at NNLO with
  McMule}}, \href{https://doi.org/10.21468/SciPostPhys.9.2.027}{\emph{SciPost
  Phys.} {\bfseries 9} (2020) 027}
  [\href{https://arxiv.org/abs/2007.01654}{{\ttfamily 2007.01654}}].

\bibitem{Engel:2019nfw}
T.~Engel, A.~Signer and Y.~Ulrich, \emph{{A subtraction scheme for massive
  QED}}, \href{https://doi.org/10.1007/JHEP01(2020)085}{\emph{JHEP} {\bfseries
  01} (2020) 085} [\href{https://arxiv.org/abs/1909.10244}{{\ttfamily
  1909.10244}}].

\bibitem{Frixione:1995ms}
S.~Frixione, Z.~Kunszt and A.~Signer, \emph{{Three-jet cross sections to
  next-to-leading order}},
  \href{https://doi.org/10.1016/0550-3213(96)00110-1}{\emph{Nucl. Phys. B}
  {\bfseries 467} (1996) 399}
  [\href{https://arxiv.org/abs/hep-ph/9512328}{{\ttfamily hep-ph/9512328}}].

\bibitem{Frederix:2009Yq}
R.~Frederix, S.~Frixione, F.~Maltoni and T.~Stelzer, \emph{{Automation of
  next-to-leading order computations in QCD: The FKS subtraction}},
  \href{https://doi.org/10.1088/1126-6708/2009/10/003}{\emph{JHEP} {\bfseries
  10} (2009) 003} [\href{https://arxiv.org/abs/0908.4272}{{\ttfamily
  0908.4272}}].

\bibitem{YENNIE1961379}
D.~Yennie, S.~Frautschi and H.~Suura, \emph{The infrared divergence phenomena
  and high-energy processes},
  \href{https://doi.org/https://doi.org/10.1016/0003-4916(61)90151-8}{\emph{Annals
  of Physics} {\bfseries 13} (1961) 379}.

\bibitem{Lepage:1980jk}
G.~P. Lepage, \emph{{VEGAS}: An adaptive multidimensional integration program}.

\bibitem{website}
{\scshape \sc McMule} collaboration, ``{The {\sc McMule} User Library}.''
  \url{https://mule-tools.gitlab.io/user-library/bhabha-scattering/validation/soft\_limit.html}.

\bibitem{Low:1958sn}
F.~E. Low, \emph{{Bremsstrahlung of very low-energy quanta in elementary
  particle collisions}},
  \href{https://doi.org/10.1103/PhysRev.110.974}{\emph{Phys. Rev.} {\bfseries
  110} (1958) 974}.

\bibitem{Burnett:1967km}
T.~H. Burnett and N.~M. Kroll, \emph{{Extension of the low soft photon
  theorem}}, \href{https://doi.org/10.1103/PhysRevLett.20.86}{\emph{Phys. Rev.
  Lett.} {\bfseries 20} (1968) 86}.

\bibitem{Beneke:1997zp}
M.~Beneke and V.~A. Smirnov, \emph{{Asymptotic expansion of Feynman integrals
  near threshold}},
  \href{https://doi.org/10.1016/S0550-3213(98)00138-2}{\emph{Nucl. Phys. B}
  {\bfseries 522} (1998) 321}
  [\href{https://arxiv.org/abs/hep-ph/9711391}{{\ttfamily hep-ph/9711391}}].

\bibitem{Smirnov:1999bza}
V.~A. Smirnov, \emph{{Problems of the strategy of regions}},
  \href{https://doi.org/10.1016/S0370-2693(99)01061-8}{\emph{Phys. Lett. B}
  {\bfseries 465} (1999) 226}
  [\href{https://arxiv.org/abs/hep-ph/9907471}{{\ttfamily hep-ph/9907471}}].

\bibitem{Jantzen:2012mw}
B.~Jantzen, A.~V. Smirnov and V.~A. Smirnov, \emph{{Expansion by regions:
  revealing potential and Glauber regions automatically}},
  \href{https://doi.org/10.1140/epjc/s10052-012-2139-2}{\emph{Eur. Phys. J. C}
  {\bfseries 72} (2012) 2139}
  [\href{https://arxiv.org/abs/1206.0546}{{\ttfamily 1206.0546}}].

\bibitem{Smirnov:1999gc}
V.~A. Smirnov, \emph{{Analytical result for dimensionally regularized massless
  on shell double box}},
  \href{https://doi.org/10.1016/S0370-2693(99)00777-7}{\emph{Phys. Lett.}
  {\bfseries B460} (1999) 397}
  [\href{https://arxiv.org/abs/hep-ph/9905323}{{\ttfamily hep-ph/9905323}}].

\bibitem{Tausk:1999vh}
J.~B. Tausk, \emph{{Nonplanar massless two loop Feynman diagrams with four
  on-shell legs}},
  \href{https://doi.org/10.1016/S0370-2693(99)01277-0}{\emph{Phys. Lett.}
  {\bfseries B469} (1999) 225}
  [\href{https://arxiv.org/abs/hep-ph/9909506}{{\ttfamily hep-ph/9909506}}].

\bibitem{Smirnov:2009up}
A.~V. Smirnov and V.~A. Smirnov, \emph{{On the Resolution of Singularities of
  Multiple Mellin-Barnes Integrals}},
  \href{https://doi.org/10.1140/epjc/s10052-009-1039-6}{\emph{Eur. Phys. J. C}
  {\bfseries 62} (2009) 445} [\href{https://arxiv.org/abs/0901.0386}{{\ttfamily
  0901.0386}}].

\bibitem{Carlo}
C.~M. Carloni~Calame, \emph{{private communication}}.

\bibitem{Gasparian:2020hog}
{\scshape PRad} collaboration, A.~Gasparian et~al., \emph{{PRad-II: A New
  Upgraded High Precision Measurement of the Proton Charge Radius}},
  \href{https://arxiv.org/abs/2009.10510}{{\ttfamily 2009.10510}}.

\bibitem{Banerjee:2021qvi}
P.~Banerjee, T.~Engel, N.~Schalch, A.~Signer and Y.~Ulrich, \emph{{M\o{}ller
  scattering at NNLO}},  \href{https://arxiv.org/abs/2107.12311}{{\ttfamily
  2107.12311}}.

\bibitem{Abbiendi:2016xup}
G.~Abbiendi et~al., \emph{{Measuring the leading hadronic contribution to the
  muon g-2 via $\mu e$ scattering}},
  \href{https://doi.org/10.1140/epjc/s10052-017-4633-z}{\emph{Eur. Phys. J. C}
  {\bfseries 77} (2017) 139}
  [\href{https://arxiv.org/abs/1609.08987}{{\ttfamily 1609.08987}}].

\bibitem{Banerjee:2020tdt}
P.~Banerjee et~al., \emph{{Theory for muon-electron scattering @ 10 ppm: A
  report of the MUonE theory initiative}},
  \href{https://doi.org/10.1140/epjc/s10052-020-8138-9}{\emph{Eur. Phys. J. C}
  {\bfseries 80} (2020) 591}
  [\href{https://arxiv.org/abs/2004.13663}{{\ttfamily 2004.13663}}].

\end{thebibliography}\endgroup

\end{document}